\documentclass[twoside,twocolumn]{article}
\usepackage[super,sort&compress,comma]{natbib}
\usepackage[version=4]{mhchem}
\usepackage{times,mathptmx}
\usepackage{sectsty}
\usepackage{balance}
\usepackage{xspace}
\usepackage{color}
\usepackage{tikz}
\usetikzlibrary{positioning}
\usepackage{multirow}
\usepackage[title]{appendix}

\usepackage{graphicx} 
\usepackage{lastpage}
\usepackage[format=plain,justification=raggedright,singlelinecheck=false,font=small,labelfont=bf,labelsep=space]{caption}
\usepackage{fancyhdr}
\usepackage{amsfonts}

\usepackage{sectsty}
\usepackage{graphicx} 
\usepackage{colortbl,booktabs}
\usepackage{array}
\usepackage{amssymb}
\usepackage{threeparttable}
\usepackage{amsmath, bm}
\usepackage{algorithm,algorithmic}

\begin{document}

\thispagestyle{plain} \fancypagestyle{plain}{
\renewcommand{\headrulewidth}{1pt}}
\renewcommand{\thefootnote}{\fnsymbol{footnote}}
\renewcommand\footnoterule{\vspace*{1pt}%
\hrule width 3.4in height 0.4pt \vspace*{5pt}}
\setcounter{secnumdepth}{5}

\makeatletter
\def\subsubsection{\@startsection{subsubsection}{3}{10pt}{-1.25ex plus -1ex minus
-.1ex}{0ex plus 0ex}{\normalsize\bf}}
\def\paragraph{\@startsection{paragraph}{4}{10pt}{-1.25ex plus -1ex minus -.1ex}{0ex
plus 0ex}{\normalsize\textit}}
\renewcommand\@biblabel[1]{#1}
\renewcommand\@makefntext[1]
{\noindent\makebox[0pt][r]{\@thefnmark\,}#1} \makeatother
\renewcommand{\figurename}{\small{Fig.}~}
\sectionfont{\large}
\subsectionfont{\normalsize}

\fancyfoot{}
\fancyfoot[RO]{\footnotesize{\sffamily{1--\pageref{LastPage}
~\textbar  \hspace{2pt}\thepage}}}
\fancyfoot[LE]{\footnotesize{\sffamily{\thepage~\textbar\hspace{3.45cm}
1--\pageref{LastPage}}}} \fancyhead{}
\renewcommand{\headrulewidth}{1pt}
\renewcommand{\footrulewidth}{1pt}
\setlength{\arrayrulewidth}{1pt} \setlength{\columnsep}{6.5mm}
\setlength\bibsep{1pt}

\twocolumn[\begin{@twocolumnfalse}
\noindent\LARGE{\textbf{Interpolative separable density fitting decomposition for accelerating Hartree-Fock exchange calculations within numerical atomic orbitals}}
\vspace{0.6cm}

\noindent\large{\textbf{Xinming Qin\textit{$^{a}$}, Jie Liu\textit{$^{a}$}, Wei Hu$^{\ast}$\textit{$^{a}$},
and Jinlong Yang$^{\ast}$\textit{$^{a}$}}}\vspace{0.5cm}

\noindent\textit{\small{\textbf{Received Xth XXXXXXXXXX 20XX,
Accepted Xth XXXXXXXXX 20XX\newline First published on the web Xth
XXXXXXXXXX 200X}}}

\noindent \textbf{\small{DOI: 00.0000/00000000}} \vspace{0.6cm}

\noindent \normalsize{
The high cost associated with the evaluation of Hartree-Fock exchange (HFX) makes hybrid functionals computationally challenging for large systems. 
In this work, we present an efficient way to accelerate HFX calculations with numerical atomic basis sets. Our approach is based on the recently proposed interpolative separable density fitting (ISDF) decomposition to construct a low rank approximation of HFX matrix, which avoids explicit calculations of the electron repulsion integrals (ERIs) and significantly reduces the computational cost. 
We implement the ISDF method for hybrid functional (PBE0) calculations in the HONPAS package. We take benzene and polycyclic aromatic hydrocarbons molecules as examples and demonstrate that hybrid functionals with ISDF yields quite promising results at a significantly reduced computational cost. Especially, the ISDF approach reduces the total cost for evaluating HFX matrix by nearly 2 orders of magnitude compared to conventional approaches of direct evaluation of ERIs.
 }
\vspace{0.5cm}\end{@twocolumnfalse} ]

\footnotetext{\textit{$^{a}$Hefei National Laboratory for Physical Sciences at the Microscale, Department of Chemical Physics, and Synergetic Innovation Center of Quantum Information and Quantum Physics, University of Science and Technology of China, Hefei, Anhui 230026, China}}
\footnotetext{\textit{E-mail: whuustc@ustc.edu.cn (Wei Hu)}} 
\footnotetext{\textit{E-mail: jlyang@ustc.edu.cn (Jinlong Yang)}}

\section{Introduction} \label{sec:Introduction}
Kohn-Sham density functional theory (DFT)\cite{Hohenberg_1964,Kohn_1965} has become one of the most popular methods for electronic structure calculations in molecules and solids due to its good balance between accuracy and computational efficiency. Despite the tremendous success, traditional local or semilocal functionals fail to describe some important properties of materials. A well-known example is that they always severely underestimate the band gaps of semiconductors. Hybrid functionals, such as B3LYP,\cite{Stephens_1994}
PBE0,\cite{Ernzerhof_1999,Adamo_1999} and HSE06,\cite{Krukau_2006} including a certain amount of Hartree-Fock exchange (HFX), can systematically improve the results of local and semilocal density functionals. However, the high computational cost of HFX limits hybrid functional calculations applying for large systems. Therefore, it is a challenging problem to develop efficient approaches for HFX calculations. 

Much effort has been devoted to improving the computational efficiency of HFX with Gaussian-type orbitals (GTOs) in the quantum chemistry community, including integral-evaluation schemes,\cite{Obara_1986,Head_Gordon_1988} integral screening approaches,\cite{Haser_1989,Lambrecht_2005,Izmaylov_2006} and linear-scaling methods.\cite{Burant_1996,Schwegler_1997,Ochsenfeld_1998,Schwegler_2000} However, it  is still computationally too expensive to construct HFX matrix through directly evaluating two-electron repulsion integrals (ERIs) with the high-quality basis set. In order to reduce the computational cost of ERIs, various alternative approximate integral-evaluation schemes, such as the Cholesky decomposition approximation,\cite{Beebe_1977,Roeggen_1986} the density fitting (DF) or resolution of the identity (RI) method,\cite{Whitten_1973,Dunlap_1979,Vahtras_1993,Weigend_2002} 
the pseudospectral (PS) approximation,\cite{Friesner_1985,Friesner_1987}
and the tensor hypercontraction (THC) approach,\cite{THC_2012,Parrish_2012,Hohenstein_2012} have been proposed. Based on these approximate schemes, linear-scaling methods have been also derived to reduce
the prefactor of HFX calculations. For example, Sodt {\em et al}. proposed the atomic resolution-of-identity for exchange (ARI-K) by employing local fitting domains,\cite{Sodt_2008} and Merlot {\em et al}. introduced a
simpler pair-atomic resolution-of-identity (PARI-K) approximation.\cite{Merlot_2013} In addition, Neese {\em et al.} developed a chain-of-spheres exchange (COSX) method by combining a semi-numeric integration with RI.\cite{Neese_2009,Izsak_2011} 

Compared to analytical GTOs, the strictly localized numerical atomic orbitals (NAOs) are more convenient and flexible for linear-scaling DFT calculations, which have been widely adopted in linear-scaling DFT codes, such as SIESTA,\cite{SIESTA_2002} CONQUEST,\cite{CONQUEST_2008} OPENMX,\cite{OPENMX_2005} FHI-aims,\cite{FHI-aims_2009} HONPAS,\cite{Qin_2015} and ABACUS.\cite{Li_2016} However, hybrid functional calculations with NAOs are rarely available since multi-center NAO integrals are fairly troublesome. Several direct schemes to calculate ERIs with NAOs have been developed,\cite{Talman_2003,Talman_2007,Toyoda_2009,Toyoda_2010,Shang_2010} but they are typically more expensive than the Gaussian-expansion methods. To overcome this problem, our group has proposed a NAO2GTO scheme to approximately evaluate ERIs by using auxiliary GTOs to represent NAOs in the HONPAS package.\cite{Shang_2011,Qin_2015} After using integral screening techniques, the HFX calculation is found to be efficient and linear-scaling. Furthermore, Ren {\em et al}. have successfully extended the RI approach to NAOs in the FHI-aims code.\cite{Ren_2012} They demonstrated that, in conjunction with a {\em priori} integral screening, the localized RI (LRI) approximation can further reduce memory consumption and lead to a linear-scaling HFX calculation.\cite{Levchenko_2015} Most recently, they have also demonstrated that the LRI approximation can provide adequate
accuracy for periodic hybrid functional calculations with NAOs in the ABACUS code.\cite{Lin_2020} Despite these developments, new approaches for efficient HFX calculations with NAOs are still urgently desired.


Recently, Lu and Ying introduced a novel algorithm called interpolative separable density fitting (ISDF)\cite{ISDF_2015} decomposition to accelerate the ERI calculations. The ISDF method has many conceptual similarities to the THC approach,\cite{THC_2012,Parrish_2012,Hohenstein_2012} but it takes a completely different strategy to construct the compressed representation of ERIs. That is, for pair products of orbital functions represented on a real space grid, ISDF uses the column-pivoted QR (QRCP) decomposition\cite{JCTC_Hu_2017} or the much simpler centroidal Voronoi tessellation (CVT) procedure\cite{CVT_2018} to select a set of interpolation points, so that the values of the orbital-pair products evaluated at such points can be used to accurately interpolate those evaluated at all grid points. Because of this smart treatment, the ISDF method is general and suitable for different atomic orbital and even canonical molecular orbital representation without specifying in advance the form of auxiliary basis functions. Then, it was applied to accelerate a number of applications, including plane-wave and Gaussian basis sets, such as hybrid density functionals,\cite{JCTC_Hu_2017,CVT_2018} random phase approximation,\cite{RPA_2017} density functional perturbation theory,\cite{DFPT_2017} linear-response time-dependent density functional theory,\cite{Hu_2020} Bethe-Salpeter equation,\cite{BSE_2018} quantum Monte Carlo simulations,\cite{JCTC_15_256_2019} and M{\o}ller-Plesset perturbation theory.\cite{MP2_2020} However, the ISDF method has not been covered yet in the context of NAOs.

In this work, we present an efficient scheme to reduce the computational cost of HFX with localized numerical atomic basis sets. Our approach is to construct a straightforward low rank approximation of HFX based on the ISDF decomposition. This approach reduces the complexity of the HFX construction with a very small prefactor. We implement the ISDF method for PBE0 calculations in the HONPAS package. We compare the ISDF method with conventional approaches, and demonstrate its performance by examining the accuracy and time of the HFX calculations in molecules of benzene and polycyclic aromatic hydrocarbons. 

The remainder of this article is organized as follows: Section~\ref{sec:Methodology} gives a brief description of the theoretical methodology, covering the HF method, the ISDF method, and their combination. Section~\ref{sec:Results} benchmarks the computational accuracy and efficiency of the ISDF decomposition to accelerate the HFX calculations. A summary and outlook is given in Section~\ref{sec:Conclusion}.

\section{Methodology} \label{sec:Methodology}

\subsection{Exact Hartree-Fock exchange (HFX)}\label{sec:hybrid_DFT}

Hybrid functionals currently used in the framework of DFT contain a fraction of nonlocal, exact HFX term. For the PBE0 functional,\cite{Ernzerhof_1999,Adamo_1999} the
exchange-correlation energy $E^\textrm{PBE0}_\textrm{xc}$ is given by
\begin{equation} \label{equ:PBE0}
E^\textrm{PBE0}_\textrm{xc}(\rho,\psi) = \dfrac{1}{4}E^\textrm{HF}_\textrm{x}(\psi) +
\dfrac{3}{4}E^\textrm{PBE}_\textrm{x}(\rho) + E^\textrm{PBE}_\textrm{c}(\rho)
\end{equation}
where $E^\textrm{HF}_\textrm{x}$ denotes the exact HFX energy, $E^\textrm{PBE}_\textrm{x}$ is the PBE exchange energy, and $E^\textrm{PBE}_\textrm{c}$ is the PBE correlation energy. For closed-shell systems, the HFX energy has an explicit expression:
\begin{equation} \label{equ:EXX_MO}
E^\textrm{HF}_\textrm{x} =
-2\sum^{N_\textrm{occ}}_{ij}{\int\int}\dfrac{\psi_{i}(\mathbf{r})\psi_{j}(\mathbf{r'})\psi_{j}(\mathbf{r})\psi_{i}(\mathbf{r'})}{|\mathbf{r}-\mathbf{r'}|}d\mathbf{r}d\mathbf{r'}
\end{equation}
where $\psi_{i}(\mathbf{r})$ are the one-electron KS orbitals. Throughout this section, the atomic units ( $\hbar= m_{e} = e = 1$) are used. Accordingly, the HFX
operator is defined by its action on an occupied orbital as
\begin{equation}   \label{eq:HFX_MO}
\hat{v}^\textrm{HFX} (\mathbf{r},\mathbf{r'})\psi_{i}(\mathbf{r'}) = -
\sum^{N_\textrm{occ}}_{j}\psi_{j}(\mathbf{r})\int\dfrac{\psi_{j}(\mathbf{r'})\psi_{i}(\mathbf{r'})}{|\mathbf{r}-\mathbf{r'}|}d\mathbf{r'}
\end{equation}

In the LCAO method, the KS orbitals are expanded as linear combinations of a set of
atomic centered basis set $\{\phi_\mu(r)\}$
\begin{equation} \label{eq:LCAO}
\psi_{i}(\mathbf{r})=
\sum^{N_\textrm{b}}_{\mu}\phi_{\mu}(\mathbf{r}) c_{\mu i}
\end{equation}
with $c_{\mu{i}}$ is the expansion coefficient at the $\mu$-th atomic orbital,
$N_\textrm{b}$ is the basis set size. Inserting Eq~(\ref{eq:LCAO}) to Eq~(\ref{eq:HFX_MO}), then the HFX energy can be expressed as:
\begin{equation}
E^\textrm{HFX}_\textrm{x}
=-\dfrac{1}{2}\sum^{N_\textrm{b}}_{\mu\nu\lambda\sigma}D_{\mu\lambda}D_{\nu\sigma}(\mu\nu|\lambda\sigma)
\end{equation}
where $D_{\mu\nu}$ are the density matrix elements
\begin{equation} \label{eq:DM}
D_{\mu\nu} = \sum^{N_\textrm{occ}}_ic_{\mu i}n_{i}c_{i\nu}
\end{equation}
with $n_i =2$ is the occupation of state $\psi_{i}$, and $(\mu\nu|\lambda\sigma)$ are the electron repulsion integrals (ERIs) defined on atomic centered orbitals
\begin{equation} \label{eq:ERI}
(\mu\nu|\lambda\sigma)={\int\int}\dfrac{\phi_\mu(\mathbf{r})\phi_\nu(\mathbf{r})\phi_\lambda(\mathbf{r'})\phi_\sigma(\mathbf{r'})}{|\mathbf{r}-\mathbf{r'}|}d\mathbf{r}d\mathbf{r'}
\end{equation}
Then, the explicit HFX matrix elements, defined as the integrations of the HFX operator $\hat{v}^\textrm{HF}_\textrm{x}$ with two atomic orbitals, are given by
\begin{equation} \label{eq:HFX-AO}
V^\textrm{HFX}_{\mu\lambda} =-\frac{1}{2}\sum_{\nu\sigma}D_{\nu\sigma}(\mu\nu|\lambda\sigma)
\end{equation}
With this representation, four-index ERIs need to be precalculated and stored first in a nondirect self-consistent field (SCF) scheme, which formally requires $\mathcal{O}(N_\textrm{b}^4)$ cost of both computation and storage, and is the major bottleneck for hybrid functional calculations. 


In our previous implementation of HFX with NAOs,\cite{Shang_2010} we have proposed a numerical scheme to calculate the ERIs by solving $N^2_\textrm{b}$ Poisson's equations for each orbital-pair products: 
\begin{equation} \label{eq:V-ISF}
V_{\mu\nu} (\mathbf{r'}) = \int \dfrac{\phi_\mu(\mathbf{r})\phi_\nu(\mathbf{r})}{|\mathbf{r}-\mathbf{r'}|}d\mathbf{r}
\end{equation}
and evaluating $N^4_\textrm{b}$ integrals in the real-space grid:
\begin{equation} \label{eq:ERI-ISF}
(\mu\nu|\lambda\sigma) = \int V_{\mu\nu}(\mathbf{r'}) \phi_\lambda(\mathbf{r'})\phi_\sigma(\mathbf{r'})d\mathbf{r'}
\end{equation}
where most of the time is spent on solving the Poisson’s equations. We employ the interpolating scaling functions (ISF) method\cite{ISF_2006} for Poisson solvers with free boundary condition, which requires a computational cost of $\mathcal{O}(N_\textrm{b}^2N_r{\log}N_r)$. Here, $N_r$ is the number of grid points in real space. Then, the grid integrals for Eq~(\ref{eq:ERI-ISF}) are straightforward and require $\mathcal{O}(N_\textrm{b}^4N_r)$ operations. Since the number of ERIs scales formally as $\mathcal{O}(N_\textrm{b}^4)$, storing and retrieving ERIs are also extremely expensive. 

In fact, the ERIs possess an eight-fold permutational symmetry. For the following exchanges of indices: $\mu \leftrightarrow \nu$, $\lambda \leftrightarrow \sigma$, and $\mu\nu \leftrightarrow \lambda\sigma$, an ERI $(\mu\nu|\lambda\sigma)$ is invariant. By utilizing the full permutational symmetry of ERIs, the number of orbital-pair products for Eq.(\ref{eq:V-ISF}) can be reduced to $N_{\mu\nu}=N_\textrm{b}(N_\textrm{b}+1)/2$, and the integral numbers for Eq.(~\ref{eq:ERI-ISF}) becomes $N_{\mu\nu}(N_{\mu\nu}+1)/2$, which leads to a factor about 8 saving in the number of ERIs to calculate and to store. The pseudocode for the standard HFX calculations is shown in Algorithm~\ref{alg:ISFforHFX}.

\begin{algorithm}[H]
\caption{The standard method for the HFX calculations.}
\leftline{\textbf{Before SCF-iterations:}}
\begin{algorithmic}[1]
\REQUIRE $\phi(\textbf{r})$ on real-space grid
\FOR{$\mu=1, N_\textrm{b}$} 
\FOR{$\nu=1, \mu$} 
\STATE{Calculate $V_{\mu\nu}$ with the ISF method  $\to $ $\mathcal{O}(N^2_\textrm{b}N_r\log N_r)$}
\FOR{$\lambda=1, \mu$} 
\STATE{$\textbf{if}  \ \  \lambda =\mu:  \sigma_{\max} = \nu $}
\STATE{$\textbf{else} \sigma_{\max} = \lambda $}
\FOR{$\sigma=1, \sigma_{\max}$} 
\STATE{Calculate $(\mu\nu|\lambda\sigma)$ by grid integrals $\to $ $\mathcal{O}(N^4_\textrm{b}N_r)$ }
\STATE{Store $\mu,\nu,\lambda,\sigma$, and $(\mu\nu|\lambda\sigma)$ in memory $\to $ $\mathcal{M}(N^4_\textrm{b})$ }
\ENDFOR
\ENDFOR
\ENDFOR
\ENDFOR
\end{algorithmic}

\leftline{\textbf{In each SCF iteration:}}
\begin{algorithmic}[1]
\REQUIRE Density matrix $D_{\nu\sigma}$ and ERIs $(\mu\nu|\lambda\sigma)$
\STATE{Read $\mu,\nu,\lambda,\sigma$, and $(\mu\nu|\lambda\sigma)$ from Memory/Disk}
\STATE{Calculate or update $V^\textrm{HFX}_{\mu\lambda}$ as Eq.~(\ref{eq:HFX-AO})  $\to $ $\mathcal{O}(N^4_\textrm{b})$ }
\end{algorithmic}
\label{alg:ISFforHFX}
\end{algorithm}

Furthermore, we can improve the computational efficiency of HFX by building an integral screening procedure during the four-index ($\mu,\nu,\lambda$ and $\sigma$) cycle. Screening approaches actually exploit the decay (sparsity) of ERIs with local basis functions, which reduces the computational cost by neglecting integrals less an easily computable upper bound.\cite{Haser_1989,Lambrecht_2005,Izmaylov_2006} In the case of NAOs, integral screening can be easily implemented by considering the strict locality of NAOs. We just need to solve $N_\textrm{screen}$ Poisson’s equations for orbital pairs that $\phi_\mu$ and $\phi_\nu$ overlap, and then integrate if $\phi_\lambda$ overlaps with $\phi_\sigma$, where $N_\textrm{screen}< N_{\mu\nu}$ is the number of orbital pairs that overlap each other. After screening, the total number of ERIs to be considered is reduced to $\mathcal{O}(N^2_\textrm{screen})$. For sufficiently large systems, $N_\textrm{screen}$ will depend linearly on $N_\textrm{b}$.


\subsection{Interpolative separable density fitting (ISDF)}
\label{sec:ISDF}

To reduce the computational cost of ERIs for constructing the HFX matrix, an alternative method is to seek a low rank representation of the fourth-order ERI
tensor. The most well-known way to achieve this is by means of the density fitting
(DF) or resolution-of-the-identity (RI) approximation,\cite{Whitten_1973,Dunlap_1979,Vahtras_1993,Weigend_2002} which exploits the fact that the highly redundant pair products
$\{\phi_\mu(\mathbf{r}) \phi_\nu(\mathbf{r})\}_{1 \leq \mu,\nu \leq N_\textrm{b}}$ in
real space can be approximately represented with a set of auxiliary basis functions (ABFs):
\begin{equation} \label{eq:RI}
\rho_{\mu\nu}(\mathbf{r})=\phi_\mu(\mathbf{r}) \phi_\nu(\mathbf{r}) \approx
\sum^{N_\textrm{aux}}_p\xi_p(\mathbf{r})C^p_{\mu\nu}
\end{equation}
where the ABFs $\{\xi_p(\mathbf{r})\}_{1\leq p \leq N_\textrm{aux}}$ are inputs that
generated either explicitly or implicitly from the original basis functions in
advance, $N_\textrm{aux}$ labels the number of ABFs and depends linearly on the original basis
size $N_\textrm{b}$, and $C^p_{\mu\nu}$ are the expansion coefficients often determined by least-squares fitting with respect to the Coulomb metric. In general, the standard DF/RI approximation can not directly reduce computational scaling of HFX due to the coupling of the indices $\mu$ and $\nu$ (and similarly $\lambda$ and $\sigma$) in the coefficients.\cite{Weigend_2002,Ren_2012} An improvement over DF/RI approximation is the grid-based tensor hypercontraction (THC) approach\cite{THC_2012,Parrish_2012,Hohenstein_2012}, which provides full separability of the four indices in the ERI tensor. However, THC often introduces additional complexity during constructing THC-ERI approximation.

The interpolative separable density fitting (ISDF) approach proposed by Lu and Ying\cite{ISDF_2015} is to compress the ERI into THC-like format. The essential idea behind the ISDF decomposition is also to take advantage of the numerically low rank nature of the orbital-pair products in real space, but it aims at the following compression
format:
\begin{equation} \label{eq:ISDF}
\rho_{\mu\nu}(\mathbf{r})=\phi_\mu(\mathbf{r}) \phi_\nu(\mathbf{r}) \approx
\sum^{N_\textrm{aux}}_p \xi_p(\mathbf{r}) \phi_\mu(\mathbf{\hat{r}}_p)
\phi_\nu(\mathbf{\hat{r}}_p)
\end{equation}
where $\phi_\mu(\mathbf{r})$ are the orbital functions discretized on a dense real space grid $\{\mathbf{r}_i\}^{N_\textrm{r}}_{i=1}$, $\{\mathbf{\hat{r}}_p\}^{N_\textrm{aux}}_{p=1}$ denote a set of interpolation points as a subset of the dense grid points, $\xi_p(\mathbf{r})$ are the corresponding interpolation vectors. Here, the number of interpolation points $N_\textrm{aux} = tN_\textrm{b}$, and $t$ is referred as the rank truncation parameter, which is the single tunable parameter to control the tradeoff between accuracy and cost. Eq.(~\ref{eq:ISDF}) states that ISDF is actually to select a set of interpolation points to interpolate the orbital-pair products at all grid points, or from a view of matrix, to select a subset of ${N_r}$ rows of $\{\rho_{\mu\nu}(\mathbf{\hat{r}}_p)\}_{N_\textrm{aux}\times N^2_\textrm{b}}$ to approximate the whole matrix $\{\rho_{\mu\nu}(\mathbf{r}) \}_{N_r\times N^2_\textrm{b}}$. 
The ISDF decomposition includes two expensive steps:\cite{JCTC_Hu_2017} (1) one is to select the interpolation points (IPs) from real space grid points by using the randomized sampling QR factorization with column pivoting (QRCP) and (2) the next is to compute the interpolation vectors (IVs) through a least-squares fitting procedure. Both these two steps only require cubic computational complexity of $\mathcal{O}(N^2_\textrm{aux}N_r)$, and the maximum memory cost scales as $\mathcal{O}(N_\textrm{aux}N_r)$.  

Compared to the traditional DF/RI method, the third-tensor $\{C^p_{\mu\nu}\}$ in ISDF is further decomposed into a transposed Khatri-Rao product of two matrices
\begin{equation} \label{eq:KRprod}
    C^p_{\mu\nu} = \phi_\mu(\mathbf{\hat{r}}_p) \phi_\nu(\mathbf{\hat{r}}_p)
\end{equation}
where the coupling of indices $\mu$ and $\nu$ are fully separated, and thus allows an overall cubic computational cost and a square storage cost for constructing HFX matrix. 

In particular, the ISDF method does not require other assumptions,\cite{ISDF_2015} {\em e.g.}, the locality of original basis functions and the form of ABFs, except that the orbitals are discrete in real space, which makes it suitable for different atomic orbital and canonical KS orbital representations. If the orbitals are sufficiently smooth and discrete on a uniform real space grid with pseudopotential approximation, ISDF will be more efficient since the number of interpolation points $N_\textrm{aux}$ selected by QRCP can be expected to be much smaller. It should be noted that, most recently, by combining with a modified centroidal Voronoi tessellation (CVT) algorithm for atom-centered grids, ISDF has also been applied to accelerate the Hartree-Fock and M{\o}ller-Plesset perturbation theory calculations with all-electron Gaussian basis sets.\cite{MP2_2020} 

Within the pseudopotential framework, the NAOs can be discretized on a uniform spatial grid for calculating pure DFT Hamiltonian matrix, as done in SIESTA code.\cite{SIESTA_2002} Hence, the QRCP-based ISDF can be applied directly. 

\subsection{Low rank representation of HFX via ISDF}\label{sec:ISDFforHFX}

Applying the ISDF decomposition to the orbital products, {\em i.e.,} by inserting Eq.~(\ref{eq:ISDF}) into Eq.~(\ref{eq:ERI}), the ERI tensor becomes
\begin{equation} \label{eq:ISDF-ERI}
(\mu\nu|\lambda\sigma)\approx \sum^{N_\textrm{aux}}_{pq}\phi_\mu(\mathbf{\hat{r}}_p) \phi_\nu(\mathbf{\hat{r}}_p)M_{pq}\phi_\lambda(\mathbf{\hat{r}}_q)\phi_\sigma(\mathbf{\hat{r}}_q)
\end{equation}
where $M_{pq}$ are the projected Coulomb integrals under the ABFs defined as
\begin{equation}\label{eq:Maux}
M_{pq} = {\int\int}\dfrac{\xi_p(\mathbf{r})\xi_q(\mathbf{r'})}{|\mathbf{r}-\mathbf{r'}|}d\mathbf{r}d\mathbf{r'}
\end{equation}
Obviously, the ERI tensor is decomposed to simple product of five matrices and the integral space of the Coulomb operate is reduced from the full orbital product space ($N_r \times N^2_\textrm{b} $) to the optimal auxiliary basis space ($N_r\times N_\textrm{aux}$) at a chosen $N_\textrm{aux}$. 

Then, the HFX matrix can be written as 
\begin{equation}
\label{eq:ISDF-HFX1}
V^\textrm{HFX}_{\mu\lambda} \approx -\dfrac{1}{2} \sum^{N_\textrm{b}}_{\nu\sigma}D_{\nu\sigma}\sum^{N_\textrm{aux}}_{pq} \phi_{\mu}(\mathbf{\hat{r}}_p)\phi_{\nu}(\mathbf{\hat{r}}_p) M_{pq} \phi_{\lambda}(\mathbf{\hat{r}}_q)\phi_{\sigma}(\mathbf{\hat{r}}_q)
\end{equation}
Because all indices $\mu$,$\nu$, $\lambda$, and $\sigma$ in Eq.~(\ref{eq:ISDF-HFX1}) are fully separated, by changing the contraction ordering, we can also obtain a matrix representation for $\textbf{V}^\textrm{HFX} \in \mathbb{R}^{N_\textrm{b} \times N_\textrm{b}}$
\begin{equation} \label{eq:ISDF-HFX2}
\textbf{V}^\textrm{HFX} \approx -\dfrac{1}{2} \Phi^\textrm{T} [(\Phi \textbf{D} \Phi^\textrm{T}) \circ \mathbf{M} ] \Phi
\end{equation}
where the orbital submatrix $\Phi = \{\phi_\mu(\mathbf{\hat{r}}_p)\}_{N_\textrm{aux} \times N_\textrm{b}}$, the density matrix $\mathbf{D} \in \mathbb{R}^{N_\textrm{b} \times N_\textrm{b}}$, the auxiliary Coulomb matrix $\mathbf{M} \in \mathbb{R}^{N_\textrm{aux} \times N_\textrm{aux}}$, and $\circ$ denotes the hadamard product. This representation is beneficial for effective matrix vectorization through the use of GEMM calls in BLAS and has a maximum scale of $\mathcal{O}(N_\textrm{b}N^2_\textrm{aux}$) since all operations are simple matrix-matrix multiplications.

In this approximation, we only need to calculate and store the auxiliary Coulomb matrix $\mathbf{M}$, where the corresponding number of integrals is much smaller than that of ERIs. We obtain the auxiliary Coulomb matrix elements in a similar way to ERIs, that is to solve the following Poisson’s equations for each $\xi_p(\mathbf{r})$
\begin{equation}\label{eq:M_split}
V_p(\mathbf{r'}) = {\int}\dfrac{\xi_p(\mathbf{r})}{|\mathbf{r}-\mathbf{r'}|}d\mathbf{r}
\end{equation}
and then to integrate in the real-space grid:
\begin{equation}\label{eq:Mpq}
M_{pq}= {\int}V_p(\mathbf{r'})\xi_q(\mathbf{r'})d\mathbf{r'}
\end{equation}
Here, the key is to solve the Poisson's equations properly, which have a Coulomb singularity at $\mathbf{r} = \mathbf{r}'$. We use the ISF method\cite{ISF_2006} with a $\mathcal{O}(N_r \log N_r)$ scaling to solve the Poisson's equations in a suitable supercell, which can avoid the singularity problem by approximating $1/{r}$ in terms of Gaussian functions. For isolated systems, it has been shown that the ISF method is more efficient than other popular methods, such as fast Fourier transform, fast multipole method, and conjugate gradients.\cite{ISF_2014} Moreover, since $\mathbf{M}$ is a real symmetric matrix, only $N_\textrm{aux}(N_\textrm{aux}+1)/2$ matrix elements ($M_{pq}$ for $p \leq q$) are required to calculate. The pseudocode of the ISDF method for HFX calculations is shown in Algorithm~\ref{alg:ISDFforHFX}. 

\begin{algorithm}[!htb]
\caption{The ISDF method for HFX calculations.}

\leftline{\textbf{Before SCF-interations:}}

\begin{algorithmic}[1]
\REQUIRE $\{\phi_\mu({\mathbf{r}_i)}\}_{i=1:N_r,\mu=1:N_\textrm{b}}$ on real-space grid, $N_\textrm{aux} = tN_\textrm{b}$
\STATE Select $N_\textrm{aux}$ interpolation points $\to$ $\mathcal{O}(N^2_\textrm{aux}N_r)$ 
\STATE Calculate and store the orbital submatrix $\Phi =\{\phi_\mu(\mathbf{\hat{r}}_p)\}_{p=1:N_\textrm{aux},\mu=1:N_\textrm{b}}$ $\to$ $\mathcal{O}(N_\textrm{b}N_\textrm{aux}) $
\STATE Calculate $N_\textrm{aux}$ interpolation vectors $\xi_p(\mathbf{r})$ $\to$ $\mathcal{O}(N^2_\textrm{aux}N_r)$


\FOR{$ p =1, N_\textrm{aux}$} 
\STATE{Calculate $V_p$ with the ISF method $\to $ $\mathcal{O}(N_\textrm{aux}N_r\log N_r)$}

\FOR{$q = 1, p$}
\STATE{Calculate $M_{pq}$ by grid integrals $\to $ $\mathcal{O}(N^2_\textrm{aux}N_r)$ }
\ENDFOR
\ENDFOR
\STATE{Store the auxiliary Coulomb matrix $\mathbf{M}_{1:N_\textrm{aux},1:N_\textrm{aux}}$ in memory $\to $ $\mathcal{M}(N^2_\textrm{aux})$ }
\end{algorithmic}
\leftline{\textbf{In each SCF iteration:}}
\begin{algorithmic}[1]
\REQUIRE Matrices ${\Phi},\mathbf{D}$, and $\mathbf{M}$
\STATE{Update $\textbf{V}^\textrm{HFX} \approx -\dfrac{1}{2} \Phi^\textrm{T}   [(\Phi \textbf{D} \Phi^\textrm{T}) \circ \mathbf{M} ] \Phi $ $\to$ $\mathcal{O}(N_\textrm{b}N^2_\textrm{aux})$ }
\end{algorithmic}
\label{alg:ISDFforHFX}
\end{algorithm}

\begin{table}[!htb] \footnotesize
\renewcommand{\arraystretch}{1.5}
\caption{Comparison of the computational cost and the memory usage for the HFX calculations by using the standard, screening and the ISDF approaches.}
\label{tab:scaling}
\centering
\renewcommand\tabcolsep{3.5pt} 
\begin{threeparttable}
\begin{tabular}{ccc}
\hline \hline
 Approach  & Cost\tnote{a} & Memory    \\
 \hline
Standard & $(N^2_\textrm{b} N_r\log N_r +N^4_\textrm{b}N_r)+N^4_\textrm{b}$ &   $N^4_\textrm{b}$ \\
Screening &$(N_\textrm{screen} N_r\log N_r + N^2_\textrm{screen}N_r)+N^2_\textrm{screen}$ & $N^2_\textrm{screen}$ \\
ISDF &$N^2_\textrm{aux}N_r +N^2_\textrm{aux}N_r+(N_\textrm{aux}N_r\log N_r+ N^2_\textrm{aux}N_r)+N^2_\textrm{aux}N_\textrm{b}$ &   $N^2_\textrm{aux}$ \\
\hline \hline
\end{tabular}
\begin{tablenotes}
\footnotesize
 \item[a] The standard and screening approaches include two terms of calculating ERIs and updating HFX, and the ISDF method includes four parts of calculating IPs, IVs, and $\mathbf{M}$ as well as updating HFX. The terms in brackets represent the calculations of ERIs or $\mathbf{M}$.  
\end{tablenotes}
\end{threeparttable}
\end{table}

Table.~\ref{tab:scaling} summarizes the scaling of computational cost and memory usage for different approaches. Compared to the standard and screening approaches, the ISDF method reduces the total number of integrals ($M_{pq}$) to $\mathcal{O}(N^2_\textrm{aux})$, where $N_\textrm{aux}$ is significantly smaller than $N_\textrm{screen}$. Correspondingly, the computational cost and the memory requirement for the Coulomb matrix are reduced to $\mathcal{O}(N_\textrm{aux}N_r\log N_r+N^2_\textrm{aux}N_r)$ and $\mathcal{M}(N^2_\textrm{aux})$, respectively. As a result, the ISDF method is expected to be very feasible for large-scale hybrid-functional calculations. Notice that the ISDF decomposition introduces two additional computational cost for selecting the IPs and calculating the IVs, both of which scale as $\mathcal{O}(N^2_\textrm{aux}N_r)$.  

%


\section{Results and discussion} \label{sec:Results}

In this section, we demonstrate the computational accuracy and efficiency of the ISDF decomposition to accelerate HFX calculations. We implement the ISDF-PBE0 calculations in the HONPAS package with the NAO basis sets under the periodic boundary condition. We use the norm-conserving pseudopotentials generated by the Troullier-Martins scheme\cite{TM_1993} to represent the interaction between core and valence electrons. Pseudopotentials constructed for the PBE functional are used throughout. All our calculations reported in this work are sequentially carried out on a single core. 

\begin{figure}[!htb]
\begin{center}
\includegraphics[width=0.5\textwidth]{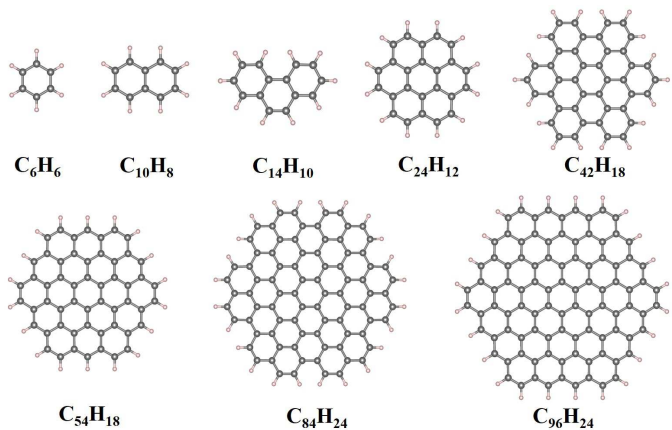}
\end{center}
\caption{Relaxed structures for benzene (C$_{6}$H$_6$) and PAHs (from C$_{10}$H$_8$ to C$_{96}$H$_{24}$).} 
\label{fig:structure}
\end{figure}

We validate the computational accuracy of the ISDF-PBE0 calculations by comparing the results with those obtained from the standard PBE0 calculations without any ERI screening. Practical tests on the accuracy are performed on two molecular systems of benzene (C$_{6}$H$_6$) and naphthalene (C$_{10}$H$_8$) with single-$\zeta$ (SZ) and double-$\zeta$ plus polarization (DZP) basis sets, respectively. A supercell of 13 {\AA} $\times$ 13 {\AA} $\times$ 8 {\AA} with a 100 Ry grid cutoff is used. For efficiency test, we choose the benzene molecule and a series of Polycyclic Aromatic Hydrocarbons (PAHs) from C$_{10}$H$_8$ to C$_{96}$H$_{24}$ with the SZ basis set. As shown in Fig.~\ref{fig:structure}, the structures of benzene and PAHs considered here are simulated in the same supercell of 25 {\AA} $\times$ 25\ {\AA} $\times$ 8\ {\AA} with a 50 Ry grid cutoff and optimized from PBE calculations. 

\subsection{Computational accuracy}\label{sec:Accuracy}

As mentioned in \ref{sec:ISDF}, the ISDF decomposition is actually a low-rank approximation of orbital product matrix $\{\rho_{\mu\nu}(\mathbf{r})\}_{N_r\times N^2_\textrm{b}}$. Thus, the computational accuracy of ISDF only depends on the number of truncated singular values of  $\{\rho_{\mu\nu}(\mathbf{r})\}$, corresponding to the number of ABFs $N_\textrm{aux}$. For tested systems here, we accurately estimate the singular values of $\{\rho_{\mu\nu}(\mathbf{r})\}$ by using exact SVD method.\cite{SVD_1970} In our tests, the truncated singular values are chosen to be $10^{-3}$, which can produce sufficiently accurate results. 

\begin{figure}[!htb]
\begin{center}
\includegraphics[width=0.5\textwidth]{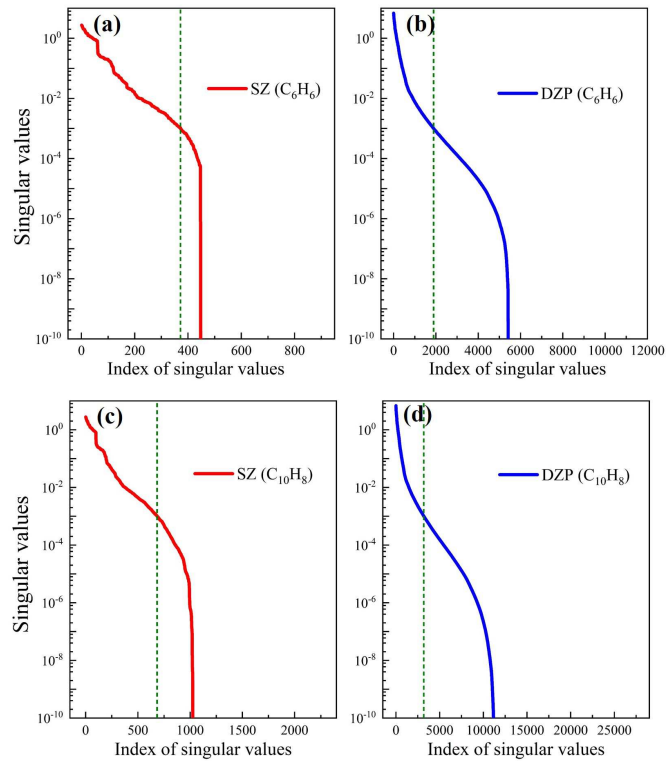}
\end{center}
\caption{ Exact singular values of the orbital product matrix for C$_6$H$_6$ and C$_{10}$H$_8$ molecules with SZ and DZP basis sets: (a) C$_6$H$_6$ SZ ($N_\textrm{b}$ = 30),  (b) C$_6$H$_6$ DZP ($N_\textrm{b}$ = 108), (c) C$_{10}$H$_8$ SZ ($N_\textrm{b}$ = 48) and  (d) C$_{10}$H$_8$ DZP ($N_\textrm{b}$ = 170). The size of the orbital product matrix is $N_r \times N^2_\textrm{b}$ with $N_r = 32000$. The dotted olive green line indicates the position where singular value is less than $10^{-3}$.}
\label{fig:SVD}
\end{figure}

\begin{table}[!htb] \footnotesize
\caption{Absolute errors in HFX energy ($\Delta E_\textrm{HFX}$ in eV/atom), total energy ($\Delta E_\textrm{tot}$ in eV/atom), and LUMO-HOMO gap ($\Delta E_\textrm{g}$ in eV) of ISDF based PBE0 calculations with varying rank parameter $t$ for C$_6$H$_6$ and C$_{10}$H$_8$ molecules with SZ and DZP basis sets. The reference results are obtained from the standard PBE0 calculations without ERI screening.}
\label{tab:accuracy}
\centering
\begin{threeparttable}
\begin{tabular}{ccccc}
\hline \hline
 $t$  &$N_\textrm{aux}/N^2_\textrm{b}$  &$\Delta E_\textrm{HFX}$ &$\Delta E_\textrm{tot}$ &$\Delta E_\textrm{g}$   \\
 \hline 
 \multicolumn{5}{c}{ C$_6$H$_6$ SZ ($N_\textrm{b} = 30$)}  \\
  \hline
6.0           &0.200      &$2.72\times 10^{-3}$        &$2.69\times 10^{-3}$    &$6.87\times 10^{-3}$      \\
8.0             &0.267      &$4.63\times 10^{-4}$        &$4.93\times 10^{-4}$    &$4.14\times 10^{-3}$     \\
10.0           &0.333      &$1.01\times 10^{-4}$        &$1.27\times 10^{-4}$    &$1.45\times 10^{-3}$     \\
12.0          &0.400      &$1.82\times 10^{-4}$        &$1.92\times 10^{-4}$    &$2.60\times 10^{-4}$     \\
14.0           &0.467      &$8.83\times 10^{-6}$        &$9.33\times 10^{-6}$    &$1.00\times 10^{-5}$   \\
 \hline
 \multicolumn{5}{c}{C$_6$H$_6$ DZP ($N_\textrm{b} = 108$)}  \\
  \hline 
6.0          &0.056     &$1.65\times 10^{-2}$      &$2.12\times 10^{-2}$    &$2.87\times 10^{-3}$   \\
8.0            &0.074     &$3.19\times 10^{-3}$      &$2.27\times 10^{-3}$    &$1.00\times 10^{-3}$  \\
10.0        &0.093     &$2.14\times 10^{-3}$      &$2.08\times 10^{-3}$   &$7.10\times 10^{-4}$   \\
12.0       &0.111     &$2.67\times 10^{-4}$      &$1.10\times 10^{-3}$    &$3.40\times 10^{-4}$  \\
14.0       &0.130     &$3.67\times 10^{-4}$      &$1.04\times 10^{-4}$    &$3.00\times 10^{-4}$  \\
16.0       &0.148     &$2.19\times 10^{-4}$      &$6.10\times 10^{-4}$    &$2.50\times 10^{-4}$  \\
18.0       &0.167     &$1.85\times 10^{-4}$      &$5.11\times 10^{-4}$    &$5.30\times 10^{-4}$ \\
20.0       &0.185     &$1.32\times 10^{-4}$       &$3.98\times 10^{-4}$    &$3.90\times 10^{-4}$ \\
24.0     &0.222     &$9.04\times 10^{-5}$       &$9.92\times 10^{-5}$    &$7.00\times 10^{-5}$ \\
\hline  
 \multicolumn{5}{c}{ C$_{10}$H$_8$ SZ ($N_\textrm{b} = 48$)}  \\
  \hline
6.0           &0.125       &$3.91 \times 10^{-3}$        &$3.43 \times 10^{-3}$    &$4.54 \times 10^{-2}$      \\
8.0         &0.167       &$2.05 \times 10^{-4}$        &$8.90 \times 10^{-5}$    &$4.75 \times 10^{-3}$     \\
10.0        &0.208       &$2.73 \times 10^{-4}$        &$3.03 \times 10^{-4}$    &$3.97 \times 10^{-3}$     \\
12.0        &0.250       &$5.38 \times 10^{-5}$        &$5.06 \times 10^{-4}$    &$1.18 \times 10^{-3}$     \\
14.0        &0.292       &$3.44 \times 10^{-5}$        &$3.32 \times 10^{-5}$    &$1.61 \times 10^{-3}$   \\
16.0       &0.333       &$1.44 \times 10^{-5}$        &$1.54 \times 10^{-5}$    &$3.60 \times 10^{-4}$     \\                                                                  
18.0         &0.375       &$1.67 \times 10^{-6}$        &$1.39 \times 10^{-6}$    &$6.00 \times 10^{-5}$   \\  
 \hline
 \multicolumn{5}{c}{C$_{10}$H$_8$ DZP ($N_\textrm{b} = 170$)}  \\       
  \hline
6.0      &0.035   &$1.82 \times 10^{-2}$      &$2.15\times 10^{-2}$    &$6.00 \times 10^{-5}$   \\
8.0       &0.047   &$1.30 \times 10^{-3}$      &$3.62\times 10^{-3}$    &$8.30 \times 10^{-4}$  \\ 
10.0       &0.059  &$1.95 \times 10^{-3}$      &$3.80\times 10^{-3}$    &$3.70 \times 10^{-4}$   \\   
12.0       &0.071  &$3.29 \times 10^{-4}$      &$4.96\times 10^{-3}$    &$4.00 \times 10^{-4}$  \\   
14.0      &0.082  &$2.57 \times 10^{-4}$      &$2.03\times 10^{-4}$    &$4.30 \times 10^{-4}$  \\   
16.0      &0.094  &$2.63 \times 10^{-4}$      &$6.48\times 10^{-4}$    &$5.30 \times 10^{-4}$  \\   
18.0      &0.106  &$2.14 \times 10^{-4}$      &$5.36\times 10^{-4}$    &$5.30 \times 10^{-4}$ \\    
20.0      &0.118  &$8.74 \times 10^{-5}$       &$3.39\times 10^{-4}$   &$3.70 \times 10^{-4}$ \\    
24.0     &0.141  &$3.03 \times 10^{-5}$      &$9.59\times 10^{-5}$    &$1.60 \times 10^{-5}$ \\    
\hline \hline
\end{tabular}
\end{threeparttable}
\end{table}

Fig.~\ref{fig:SVD} plots the decremented singular values of $\{\rho_{\mu\nu}(\mathbf{r})\}$ for benzene and naphthalene with different basis sets. We observe that the singular values decay rapidly as the index increases, especially in the case of DZP basis set.
For benzene with SZ and DZP basis sets, the numbers of dominated singular values ($> 10^{-3}$) are respectively 372 (out of 465) and 1892 (out of 5886), which corresponds to the rank truncation ratios $N_\textrm{aux}/N_{\mu\nu}$ of 0.80 and 0.32. The rank truncation is thus more significant as the size of basis set increases. Consequently, we expect no significant loss of accuracy when the rank truncation parameters $t = N_\textrm{aux}/N_\textrm{b}$ are set to 12.40 and 17.5. For larger system of naphthalene, the numbers of dominated singular values are respectively 684 (out of 1176) and 3193 (out of 14535) for SZ and DZP basis sets, the corresponding rank truncation ratios become smaller (0.58 and 0.22 ) with $t$ = 14.3 and 11.8. Previous results suggest that $t$ is independent of the system size, and $ t \approx 10.0$ can usually yield a total energy accuracy of $10^{-4}$ Hartree/atom.\cite{JCTC_Hu_2017,CVT_2018} Since $t$ obtained by truncating the singular values is roughly in this range, we also expect similar results in this work.

To verify our prediction, we measure the accuracy of ISDF-PBE0 calculations in terms of the HFX energy, the total energy of PBE0, and the energy gap between the highest occupied molecular orbital (HOMO) and the lowest unoccupied molecular orbital (LUMO). The absolute errors in the HFX energy, the total energy and the LUMO-HOMO gap are respectively defined as
\begin{equation}\label{eq:errors}
\begin{split}
&\Delta E_\textrm{HFX} = |E^\textrm{ISDF}_\textrm{HFX}-E^\textrm{exact}_\textrm{HFX}|/N_\textrm{atom}\\
&\Delta E_\textrm{tot} = |E^\textrm{ISDF}_\textrm{tot}-E^\textrm{exact}_\textrm{tot}|/N_\textrm{atom}\\
&\Delta E_\textrm{g} = |E^\textrm{ISDF}_\textrm{g}-E^\textrm{exact}_\textrm{g}|; \ E_\textrm{g} = \epsilon_\textrm{LUMO}-\epsilon_\textrm{HOMO} \\
\end{split}
\end{equation}
where $N_\textrm{atom}$ is the number of atoms, the ISDF results are obtained from ISDF-PBE0 calculations with different values of $t$, and the exact results are from the standard PBE0 calculations without ERI screening. Two molecules of benzene and naphthalene with SZ and DZP basis sets are tested to determine the effect of basis set and system size on accuracy, and the rank truncation parameter $t$ is increased from 6.0 to 24.0. As listed in Table.~\ref{tab:accuracy}, we observe that the accuracy of the ISDF-PBE0 calculations can be improved systematically by increasing the rank truncation parameter $t$. When $t$ is set to a small value of 6.0, the energy errors for all tested systems are already below the chemical accuracy of 1 kcal/mol ($4.3 \times 10^{-2}$ eV/atom). In particular, we find that a value of $ t \approx 12.0-14.0$ (12.0 for SZ and 14.0 for DZP) is sufficient to converge the total energy error bellow 1 meV/atom, which is in good agreement with previous results.\cite{JCTC_Hu_2017,CVT_2018} These results demonstrate that $t$ is roughly independent of both basis set and system sizes even if the rank truncation ratio significantly decreases (see Fig.~\ref{fig:SVD}). Therefore, the number of ABFs $N_\textrm{aux}$ scales linearly with $N_\textrm{b}$.

\begin{figure}[!htb]
\begin{center}
\includegraphics[width=0.5\textwidth]{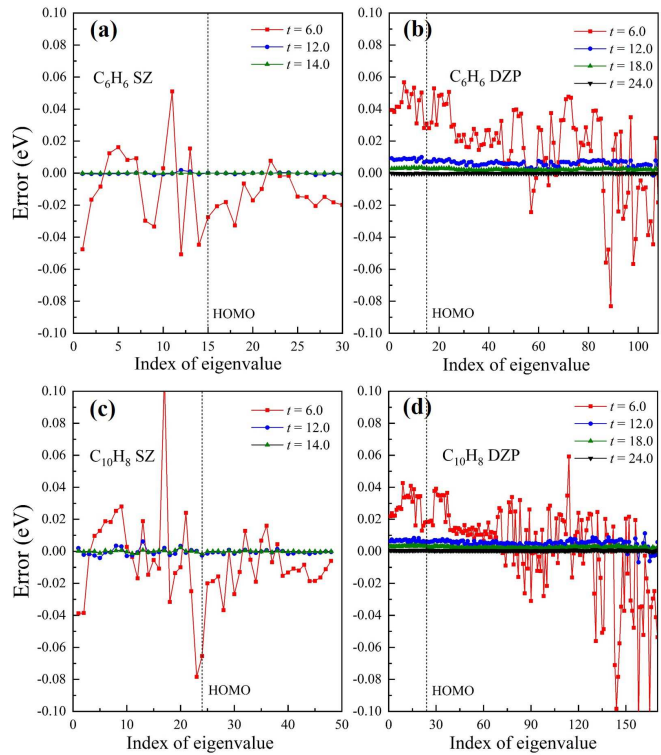}
\end{center}
\caption{The variation of eigenvalue error with respect to different values of $t$ for (a-b) benzene and (c-d) naphthalene
molecules with SZ and DZP basis stets. Three and four rank truncation parameters are used for SZ ($t$ = 6.0, 12.0, and 14.0) and DZP ($t$ = 6.0, 12.0, 18.0, and 24.0) basis sets, respectively. The dotted line represents the position of HOMO level.}
\label{fig:eigenerror}
\end{figure}

From Table.~\ref{tab:accuracy}, we can also see that the LUMO-HOMO gap error is generally less than 0.01 eV, especially for C$_{10}$H$_8$ with DZP basis set, where the corresponding error is kept under 1 meV for all considering values of $t$ here. To understand this in detail, we calculate all eigenvalue errors for the ground-state Hamiltonian matrix. The eigenvalue error is computed as $\Delta \epsilon_i = \epsilon_i^\textrm{ISDF}-\epsilon_i^\textrm{exact}$, where $i \in (1, N_\textrm{b})$ is the index of eigenvalue. Fig.~\ref{fig:eigenerror} shows the variation of eigenvalue error with respect to different values of $t$ for the benzene and naphthalene molecules with SZ and DZP basis sets. Clearly, a small rank truncation parameter $t = 6.0$ gives a significant amount of eigenvalue error ($>$ 0.04 eV) in all cases. All errors are negligible ($< 10^{-4}$ eV) only when $t$ is respectively set to 14.0 and 24.0 for SZ and DZP basis sets. Interestingly, in the case of DZP basis set, the eigenvalue errors for almost all levels are positive with the same magnitude, which indicates that the energy levels shift similarly at a given $t$. Since the HOMO and LUMO levels may introduce almost the same shift by the ISDF approximation, the LUMO-HOMO gap error would be very tiny, as shown by the results of naphthalene molecule with DZP basis set. This suggests that a relatively small rank truncation parameter can be used for fast prediction of energy gaps. It should be mentioned that, for a large basis set, the eigenvalue error of low energy levels (occupied states) is generally smaller than that of high energy levels (unoccupied states), and thus iterative diagonalization methods\cite{Davidson1975,Lanczos1988,SIAMJSC_23_517_2001_LOBPCG} to further accelerate HFX calculations by solving the projected eigenvalue problem in a small subspace ({\em e.g.}, 2-3 times of $N_\textrm{occ}$) is also expected.

\subsection{Computational efficiency}\label{sec:Efficiency}

In this section, we measure the computational efficiency and scaling behavior of the ISDF decomposition to accelerate the PBE0 calculations. Taking coronene (C$_{24}$H$_{12}$) with the SZ basis set as an example, in Table.~\ref{tab:cost} we report the total time (in s) of HFX calculations with the ISDF decomposition compared with the standard and screening approaches. Without using any approximation, the total time of standard HFX calculation is up to 34082.37 s, in which almost all the time is spent on the calculation of ERIs. Since ERI screening reduces the number of ERIs to be considered (see Table.~\ref{tab:number}), the total time can be reduced to 14638.40 s by using the screening method, but this reduction is not satisfactory. By contrast, the ISDF decomposition significantly reduces the cost of HFX calculations from hours to minutes. When the rank parameter $t$ is set to 6.0 (12.0), the total time is reduced to only 425.44 (1320.94) s. Notice that updating the HFX matrix for all approaches considered here is negligible. Therefore, the ISDF method can accelerate the PBE0 calculations by about 2 orders of magnitude compared to the standard or screening approaches.

\begin{table}[!htb] \footnotesize
\renewcommand{\arraystretch}{1.5}
\caption{Comparison of computational time (in s) spent in the HFX calculations by using the standard and screening approaches as well as the ISDF method with different $t$ for the coronene (C$_{24}$H$_{12}$) molecule with the SZ basis set.}
\label{tab:cost}
\centering
\begin{tabular}{cccccc}
\hline \hline
Approach &\multicolumn{3}{c}{ERIs}
 &HFX &Total \\
 \hline
Standard & \multicolumn{3}{c}{34082.23} &0.14 &34082.37 \\
Screening  &\multicolumn{3}{c}{14638.26} &0.139 & 14638.40 \\
\cline{2-4}
 &IPs &IVs &$\mathbf{M}$  &  & \\
 \cline{2-4}
ISDF ($t=24.0$) &2542.47 &170.91 &1945.15 &0.105 &4663.47\\
ISDF ($t=12.0$) &687.98  &58.38  &571.92 &0.021 &1320.94 \\
ISDF ($t=6.0$)  &210.99  &23.07  &189.86 &0.006 &425.44 \\
\hline\hline
\end{tabular}
\end{table}

In Table.~\ref{tab:cost} we also show the computational time of time-consuming steps for ISDF ($t$ = 6.0,  12.0, and 24.0), including selecting the interpolation points (IPs), computing the interpolation vectors (IVs) and constructing the auxiliary Coulomb matrix ($\mathbf{M}$). For all cases, selecting the IPs is always the most time-consuming step since we use the relatively expensive randomized QRCP procedure. However, this cost is expected to be further reduced by using the centroidal Voronoi tessellation (CVT) method, which is able to yield similar accuracy at a much lower computational cost. Similarly, constructing the auxiliary Coulomb matrix is almost as expensive as selecting the IPs, which can actually be improved by using corrected reciprocal FFT-based methods, especially for periodic neutral systems. Computing the IPs costs about $1/10$ of the above two steps, but it is difficult to reduce.

\begin{table}[!htb] \footnotesize
\renewcommand{\arraystretch}{1.5}
\caption{Comparison of the effective number of orbital pairs or ABFs in the standard, screening, and ISDF approaches for molecules from C$_{6}$H$_6$ to C$_{42}$H$_{18}$ with the SZ basis set.}
\label{tab:number}
\centering

\begin{tabular}{ccccccc}
\hline \hline
Approach  &C$_6$H$_6$ & C$_{10}$H$_8$ &C$_{14}$H$_{10}$ &C$_{24}$H$_{12}$  &C$_{42}$H$_{18}$ \\
\hline
Standard ($N_{\mu\nu}$) & 465  &1176  &2211  &5886 &17391 \\
Screening ($N_\textrm{screen}$) &465  &1084  &1748  &3856  &7362 \\
ISDF ($N_\textrm{aux}$)    &360  &576   &792    &1296 &2232 \\
\hline \hline
\end{tabular}
\end{table}

As discussed in Section.~\ref{sec:ISDFforHFX}, the computational cost of the standard, screening, and ISDF approaches is determined respectively by the parameters $N_{\mu\nu}$, $N_\textrm{screen}$, and $N_\textrm{aux}$, where $N_{\mu\nu} = N_\textrm{b}(N_\textrm{b}+1)/2$ is the effective number of orbital pairs after considering the permutational symmetry, $N_\textrm{screen}$ further considers the overlap of orbital pairs on $N_{\mu\nu}$, and $N_\textrm{aux} = tN_\textrm{b}$ is the number of the ABFs. Table.~\ref{tab:number} lists the corresponding values of $N_{\mu\nu}$, $N_\textrm{screen}$, and $N_\textrm{aux}$ for the molecules from C$_{10}$H$_8$ to C$_{42}$H$_{18}$ with the SZ basis set. For our tested systems, we find that $N_\textrm{aux} < N_\textrm{screen} \leq N_{\mu\nu}$, and $N_\textrm{screen}$ is significantly less than $N_{\mu\nu}$ only when the system size is large. In fact, $N_\textrm{screen}$ should also scale linearly with $N_\textrm{b}$ if we test for larger systems.  
Nevertheless, it is difficult to seek for the minimized number of integrals by ERI screening, while the ISDF decomposition can easily achieve this by adjusting rank truncation of orbital product matrix $\{\rho_{\mu\nu}(\mathbf{r})\}$. 

\begin{figure}[!htb]
\begin{center}
\includegraphics[width=0.5\textwidth]{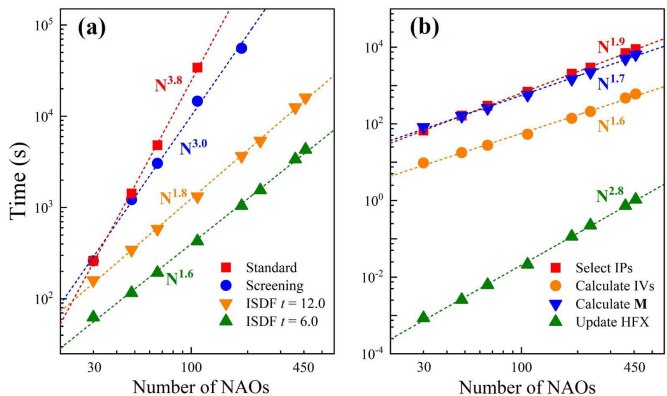}
\end{center}
\caption{ (a) Total time (in s) of HFX calculations in the standard, screening and the ISDF ($t$ = 6.0 and 12.0) methods as a function of the number of NAOs. (b) Computational time (in s) in the major steps of the ISDF-HFX method with $t$ = 12.0 as a function of the number of NAOs. The tested systems are the benzene and PAHs from C$_{10}$H$_8$ to C$_{96}$H$_{24}$ with the SZ basis set.}
\label{fig:scaling}
\end{figure}

Furthermore, since $N_\textrm{aux}$ scales linearly with $N_\textrm{b}$, the ISDF method is also expected to reduce the scaling of HFX calculation from $\mathcal{O}(N^4_\textrm{b})$ to $\mathcal{O}(N^2_\textrm{b})$ if the number of grid points $N_r$ is not considered. Using different approaches, we perform the HFX calculations for a series of molecules from C$_{6}$H$_6$ to C$_{96}$H$_{24}$ with the SZ basis set. In our test, all molecules are placed in the same supercell so that the number of grid points ($N_r$ = 419904) is fixed. The variation of total time with respect to the number of NAOs is plotted in Fig.~\ref{fig:scaling}(a). We can see that the scaling of standard HFX calculations is fitted to $\mathcal{O}(N^{3.8}_\textrm{b})$, which is close to $\mathcal{O}(N^{4}_\textrm{b})$. When the rank parameters are set to 12.0 and 6.0, the fitted scalings for ISDF-HFX calculations are $\mathcal{O}(N^{1.8}_\textrm{b})$ and $\mathcal{O}(N^{1.6}_\textrm{b})$, respectively, both of which are close to $\mathcal{O}(N^{2}_\textrm{b})$. Thus, the ISDF method shows a reduced computational scaling as predicted, in agreement with the previous results.\cite{MP2_2020} The screening approach also reduces the computational cost of HFX, but it scales as $\mathcal{O}(N^{3.0}_\textrm{b})$ for our tested systems. 

To show the scaling of ISDF-HFX calculation in detail, in Fig.~\ref{fig:scaling} (b) we present the computational time of selecting the IPs and computing the IVs in the ISDF part, constructing the auxiliary Coulomb matrix, and updating the HFX matrix as a function of the number of NAOs. As it can be observed, only the cost of updating the HFX matrix scales as $\mathcal{O}(N^{2.8}_\textrm{b})$, while the cost of all other steps has predicted scaling close to $\mathcal{O}(N^{2}_\textrm{b})$. Nevertheless, the prefactor for the calculation of updating the HFX matrix is very small and negligible, the overall scaling of ISDF-HFX calculations thus remains $\mathcal{O}(N^{2}_\textrm{b})$. Furthermore, we choose a series of molecules with planar structures as tests, in which the effect of the number of grid points $N_r$ is not included. Actually, all computational scaling except that of updating the HFX matrix should be added by 1 order since $N_r$ scales linearly with $N_\textrm{b}$.


\section{Conclusion and outlook} \label{sec:Conclusion}

In summary, we apply the interpolative separable density fitting (ISDF) decomposition to accelerate Hartree-Fock exchange (HFX) calculations within numerical atomic orbital basis sets. The ISDF method allows us to reduce the complexity of the HFX matrix construction with a small prefactor. We show that this method accurately yields the PBE0 energies in molecules of benzene and polycyclic aromatic hydrocarbons with significantly reduced computational cost. Compared to conventional approaches with and without integral screening, our implementation of ISDF reduces the total cost of HFX calculations from hours to minutes. We believe that the ISDF method may be an important trend for fast hybrid functional calculations within numerical atomic orbitals.

The performance results presented in this work are based on a sequential implementation of ISDF-PBE0, in which the efficiency test is limited by the size of the system. In the near future, we will develop a parallel ISDF implementation that allow us to deal with much larger systems. Furthermore, the ISDF approach for accelerating HFX calculations reported here remains computationally expensive in both selecting the interpolation points and constructing the auxiliary Coulomb matrix. In the future, we plan to replace the costly QRCP procedure with the centroidal
Voronoi tessellation (CVT) method\cite{CVT_2018} for selecting the interpolation points, which is expected to produce similar accuracy at a much lower computational cost. We are also exploring fast algorithms based on fast Fourier transforms for Poisson solvers with an adequate treatment of the Coulomb singularities, {\em e.g.} by using the truncated Coulomb potential\cite{Singur_2008} or auxiliary functions.\cite{Singur_2009}

\section*{Acknowledgements}
 
This work is partly supported by the National Natural Science Foundation of China (21688102, 21803066), by the Chinese Academy of Sciences Pioneer Hundred Talents Program (KJ2340000031), by the National Key Research and Development Program of China (2016YFA0200604), the Anhui Initiative in Quantum Information Technologies (AHY090400), the Strategic Priority Research Program of Chinese Academy of Sciences (XDC01040100), the Fundamental Research Funds for the Central Universities, the Research Start-Up Grants (KY2340000094) and the Academic Leading Talents Training Program (KY2340000103) from University of Science and Technology of China. The authors thank the National Supercomputing Center in Wuxi, the Supercomputing Center of Chinese Academy of Sciences, the Supercomputing Center of USTC, and Tianjin, Shanghai, and Guangzhou Supercomputing Centers for the computational resources. We would like to thank Prof. Honghui Shang (CAS), Prof. Xinguo Ren (USTC), and Prof. Lin Lin (UC Berkeley) for informative discussions and suggestions.

\bibliography{rsc}
\bibliographystyle{rsc}

\end{document}